\documentclass[aps,pra,twocolumn,longbibliography,superscriptaddress]{revtex4-1}
\usepackage{amssymb}
\usepackage{paralist}
\usepackage{verbatim}
\usepackage{chemarr}
\usepackage{braket}
\usepackage{amsmath}
\usepackage{amsthm}
\usepackage{latexsym}
\usepackage{amssymb}
\usepackage{bm}
\usepackage{bbm}
\usepackage{graphics,epstopdf}
\usepackage{physics}
\usepackage{color}
\usepackage[colorlinks=true,linkcolor=blue,citecolor=red,plainpages=false,pdfpagelabels]{hyperref}
\usepackage[normalem]{ulem}
\usepackage{newlfont}
\usepackage{amsfonts}
\usepackage{amsthm}
\usepackage{mathtools}
\usepackage{graphicx}
\usepackage{epsfig}
\usepackage{mathrsfs}
\usepackage[thinc]{esdiff}
\usepackage{caption}
\usepackage{subcaption}
\usepackage{framed}
\usepackage{times}
\usepackage{tikz}
\newcommand\scalemath[2]{\scalebox{#1}{\mbox{\ensuremath{\displaystyle #2}}}}
\newtheorem{definition}{Definition}

\begin{document}

\title{Entanglement Meter: Estimation of entanglement with single copy in Interferometer}

\author{Som Kanjilal}\email{somkanjilal@hri.res.in}
\affiliation{Harish-Chandra Research Institute,\\  A CI of Homi Bhabha National
Institute, Chhatnag Road, Jhunsi, Prayagraj 211019, India
}

\author{Vivek Pandey}\email{vivekpandey@hri.res.in}

\affiliation{Harish-Chandra Research Institute,\\  A CI of Homi Bhabha National
Institute, Chhatnag Road, Jhunsi, Prayagraj 211019, India
}
\author{Arun Kumar Pati}\email{akpati@hri.res.in}
\affiliation{Harish-Chandra Research Institute,\\  A CI of Homi Bhabha National
Institute, Chhatnag Road, Jhunsi, Prayagraj 211019, India
}
\begin{abstract}
Efficient certification and quantification of high dimensional entanglement of composite systems are challenging both theoretically as well as experimentally.
Here, we demonstrate that several entanglement detection methods can be implemented efficiently in a Mach-Zehnder Interferometric set-up. In particular, we demonstrate how to measure the linear entropy and the negativity of bipartite systems from the visibility of Mach-Zehnder interferometer using single copy of the input state. Our result shows that for any two qubit pure bipartite state, the interference visibility is a direct measure of entanglement. We also propose how to measure the mutual predictability experimentally from the intensity patterns of the interferometric set-up without having to resort to local measurements of mutually unbiased bases. Furthermore, we show that the entanglement witness operator can be measured in a interference setup and the phase shift is sensitive to the separable or entangled nature of the state. Our proposals bring out the power of Interferometric set-up in entanglement detection of pure and several mixed states which paves the way towards design of entanglement meter.
\end{abstract}

\maketitle

\section{Introduction}

Since its discovery by Einstein, Podolsky and Rosen~\cite{EPR1935}, entanglement has been firmly established as one of the most important features of quantum theory. Quantum entanglement has been helpful in some important discoveries in quantum information like quantum teleportation~\cite{Bennett1993a}, quantum dense coding~\cite{Wiesner1992}, quantum cryptography~\cite{Ekert1991}, remote state preparation~\cite{Pati2000}, where entanglement acts as a resource.  Recently, interest has been devoted towards the possibility of generating high dimensional entangled states. Such states can, in principle, contain a large amount of entanglement, which is not only conceptually interesting but also offers novel perspectives for applications in quantum information, particularly in quantum communications \cite{BS+99,BT00,CB+02,SS10,VP+10}. There are several experimental implementations of high dimensional entanglement. In particular, using photonic system, high dimensional entanglement can be created using energy time \cite{F89,ta+04,rf+12}, time bins \cite{bg+99,tt16,sg05}, orbital angular momentum \cite{mv+01,da+11,kh+14}, and frequency modes \cite{om+12,bb+13,js+16}. Also, the entanglement of these states can be detected experimentally, via the use of entanglement witnesses or Bell inequalities \cite{pc+12,bc+14}.

However, efficient experimental certification and quantification of high dimensional entanglement is still a challenging problem. There are mainly two reasons why this is demanding. Firstly, the characterization of a high dimensional entangled state via standard methods (e.g., quantum tomography) typically requires the estimation of a large number of independent parameters, which in turn requires a large number of different measurements to be performed. Furthermore, it has also been comprehensively argued that, even with full knowledge of the density matrix of the state obtained from tomography, the computational complexity of evaluating any entanglement monotone increases so rapidly with increasing dimension making it a NP-hard problem \cite{hn+17,h14}. Secondly, the types of measurements that can actually be performed in a real experiment are typically limited. 

To circumvent the above-mentioned difficulties various methods have been proposed in recent years for states with high purity \cite{gl+10,tc+14,hk+14,bg+16}. Other approaches include determination of Schmidt number \cite{ss+13,ss+14}, comparison of measurement results corresponding to two or more mutually unbiased bases \cite{gr+13,sc+12}, using the violation of the entropic EPR-Steering inequality \cite{sh+18} or other suitably defined statistical correlators \cite{ek+17}. 

Against this backdrop, we explore whether one can use Mach Zehnder interferometer to design entanglement meter where one can detect and estimate the entanglement of bipartite quantum states directly from the interference pattern. It is known that linear and non-linear functionals of the density matrix can be measured from the Mach Zehnder interferometric set-up \cite{SE+00},  where the functionals of the density operator are taken to be various measures and witnesses of entanglement. It has been recently shown that, compared to quantum state tomography, one can determine the quantum state more efficiently using a process called quantum state interferography~\cite{ss+20}. Inspired by this, we attempt to see whether one can determine entanglement of two-qubit states as well as higher dimensional states from the Mach-Zehnder interferometer \cite{SE+00,zp+15} more efficiently and more economically without quantum state tomography and classical post processing. In Ref.~\cite{zp+15}, it has been shown that the average visibility in interference is related to concurrence of the quantum state. However, the proposal described in Ref.~\cite{zp+15} involves averaging over multiple execution instances. To be precise, the required operations (random unitaries) are sampled from from a suitable set of operations to determine the amount of entanglement from the interference visibility. In this paper, we show that there exists a single unitary operation and single copy of the input state such that using the interference visibility we can determine whether the bipartite state is entangled or not and how much entanglement is there. It was shown in Refs.~\cite{Horodecki2002,EA+02,Bovino2005,Mintert2007} that the linear entropy can be measured via interference setup using two copies of the state. In this paper, we show that a single copy of the input state is sufficient to measure the linear entropy in interference setup.  
 
In addition, we also provide a scheme to determine the negativity measure of entanglement using single copy of the joint state at the input port of the interferormeter. In particular, our 
method can determine the negativity of bipartite pure state and certain mixed state from the interference visibility using a single unitary operation acting on a single copy of the state. 
Furthermore, we propose how to measure the mutual predictability directly from the intensity patterns of the interferometric set-up without having to resort to local measurements of mutually unbiased bases. We also show that the entanglement witness operator can be measured in an interference setup and the phase shift is sensitive to the separable or entangled nature of the state. This can have an interesting application in design of phase sensitive entanglement meter where one can have a portable device to detect entangled or separable states using the phase shift (similar to the concept of voltmeter which measures the voltage between two points). The notion of entanglement meter will have several applications in quantum technology as this can directly display how much entanglement is there between a pair of systems.

The structure of this paper is as follows. In Section~\ref{prelim}, we describe the interferometric set-up for mixed state introduced in Ref.~\cite{SE+00}. In Section~\ref{sub4}, we show that the linear entanglement entropy and the negativity of a pure bipartite entangled state can be obtained from the interferometric visibility and the phase shift. In Section~\ref{subA}, we provide a method for detecting entanglement via the mutual predictability using interferometric setup, for finite dimensional bipartite pure states.  In
Section~\ref{subB}, we provide a method for detecting entanglement from interferometric visibility and phase shift via an entanglement witness, namely the swap operator, for finite dimensional bipartite pure states and mixed state as well.  Finally, in Section~\ref{conclusion}, we summarise our findings.

\section{Preliminaries}
\label{prelim}
For the sake of completeness, we first describe the interference of mixed states as formalized in Ref.~\cite{SE+00}. A typical interferometric setup for a system is a sequence of of the following operations: taking the arms of the beam splitter as the basis states $\ket{0}$ and $\ket{1}$ and the incoming quantum state as $\rho$, first we apply the Hadamard gate $H$ (which corresponds to the first beam splitter), followed by a phase shift $\phi$ in one of the arms (say the arm corresponding to $\ket{0}$) , Hadamard gate $H$ (which corresponds to the final beam splitter) and finally a measurement in the computational basis. We insert a controlled-unitary operation between the Hadamard gates, with its control on the qubit and with the unitary acting on a quantum system described by some unknown density operator $\rho$. The above sequence of gates results in the evolution operator
\begin{align}
U_{tot} & = (H\otimes \mathbb{I})(\ketbra{0}{0}\otimes\mathbb{I}+\ketbra{1}{1}\otimes U)\nonumber \\
\label{evolop}
& \hspace{0.35cm} ({ \mathrm e}^{i\phi}\ketbra{0}{0}\otimes\mathbb{I}+\ketbra{1}{1}\otimes\mathbb{I})(H\otimes \mathbb{I}),
\end{align}
 where $\mathbb{I}$ is the identity. Following figure summarizes the aforementioned scheme.
\begin{framed}
\begin{center}
\begin{tikzpicture}
\node at (-0.5,0) {$\ket{0}$}; 
\draw (0,0) -- (2,0);
\draw (2,-0.5) -- (2,0.5) -- (3,0.5) -- (3,-0.5) -- (2,-0.5);
\node at (2.5,0) {$H$};
\draw (3,0)--(4,0);
\node at (4,0.5) {$\phi$};
\filldraw[black] (4,0) circle (2pt);
\draw (4,0)--(5,0);
\node at (5.5,0) {$H$};
\draw (6,0) -- (6.65,0);
\draw[black](7,0) circle (10pt);
\node at (7,0) {$M$};
\draw (5,-0.5) -- (5,0.5) -- (6,0.5) -- (6,-0.5) -- (5,-0.5);
\draw (4.5,0)--(4.5,-1.5);
\node at (4.5,-2) {$U$};
\draw (4,-1.5) -- (5,-1.5) -- (5,-2.5) -- (4,-2.5) -- (4,-1.5);
\draw (0,-2) -- (4,-2);
\draw (5,-2) -- (6,-2);
\node at (-0.5,-2) {$\rho$};

\end{tikzpicture}
\label{figure}
\end{center}
The initial state $\ket{a}$, where $a\in \{0,1\}$, converts to $\frac{1}{\sqrt{2}}(\ket{0}+(-1)^{a}\ket{1})$ after the first beam beam-splitter. This transformation is modelled using a Hadamard Gate $H$. Then a phase-shifter $\phi$ is inserted in arm $\ket{0}$. After that, a controlled-Unitary operation $U$ with interferrometric arms as control qubit is applied on $\rho$. Finally, the second beam-splitter is modelled as Hadamard Gate. Here $M$ is the measurement in the $\{\ket{0},\ket{1}\}$ basis.
\end{framed}

 If we start with a state $\ketbra{0}{0}\otimes\rho$ then the above sequence of gates give rise to the following final state
\begin{align}
\label{finalstate}
\rho_{f} &= \frac{1}{2}[\ketbra{+}{+}\otimes U\rho U^{\dagger} + \ketbra{-}{-}\otimes\rho+ { \mathrm e}^{-i\phi}\left(\ketbra{+}{-}\otimes U\rho\right) \nonumber \\
 & \hspace{0.35cm} +{ \mathrm e}^{i\phi}\left(\ketbra{-}{+}\otimes \rho U^{\dagger}\right)],
\end{align}
where $\ket{+}= \frac{1}{\sqrt{2}}\left(\ket{0}+\ket{1}\right)$ and $\ket{-}= \frac{1}{\sqrt{2}}\left(\ket{0}-\ket{1}\right)$. If we detect the intensity corresponding to the interferometric arm $\ket{0}$ it results in 
\begin{align}
\mathcal{I} & = \tr\left(\left(\ketbra{0}{0}\otimes\mathbb{I}\right)\rho_{f}\right) \nonumber\\
\label{intensity}
& = \frac{1}{4}[1+\left|\tr(U\rho)\right|\cos(\phi-\arg[\tr(U\rho)])], 
\end{align}
where $\tr(U\rho)=\left|\tr(U\rho)\right|\exp[\arg[\tr(U\rho)]]$. It is seen that the action of the controlled-unitary on $\rho$ modifies the visibility of the interference pattern by the factor $V = \left|\tr(U\rho)\right|$ and the phase is modified by the argument of $\tr(U\rho)$, i.e.,  $\alpha =\arg[\tr(U\rho)]$ for an input state $\rho$ undergoing unitary evolution \cite{SE+00}. Thus, we can infer the quantity $\tr(U\rho)$ by measuring the change in the visibility and the  phase shift. We will show that we can choose the unitary $U$ suitably, such that, from the quantity $\tr(U\rho)$, we can determine whether the state is entangled and the amount of entanglement present in the bipartite system.
In particular, the unitary $U$ can be chosen to measure the linear entanglement entropy~\cite{U00}, the negativity~\cite{VW02}, the mutual predictability~\cite{SH+12} of $\rho_{AB}$ and determine the expectation value of an entanglement witness operator for the composite state. After that, we also provide a scheme such that $|\tr(U\rho)|$ is proportional to the negativity measure of entanglement for pure state and certain kinds of mixed state. Note that in Ref.~\cite{zp+15,zz+20}, the measure of entanglement in the quantum state $\rho$ is determined by averaging over randomly distributed unitaries. In contrast, we show that only a single unitary can be used to determine the the amount of entanglement of the state.

For the rest of the article, we will consider that \(\mathcal{H}\) represent a separable Hilbert space with ${\mathrm{ dim}}(\mathcal{H})$ denoting the dimension of the Hilbert space. 
We let  \(\mathcal{H}_{A}\) and  \(\mathcal{H}_{B}\) denote the Hilbert space associated with quantum system $A$ and $B$, respectively. 
The Hilbert space of composite system $AB$ is denoted by $\mathcal{H}_{AB} = \mathcal{H}_{A}\otimes\mathcal{H}_{B}$. We denote $\ket{\Psi}_{AB}$ as pure entangled state and $\rho_{AB}$ as a mixed entangled state in $\mathcal{H}_{A}\otimes\mathcal{H}_{B}$. Consider a pure bipartite state $\ket{\Psi}_{AB} \in \mathcal{H}_{A}\otimes\mathcal{H}_{B} $, where $\mathcal{H}_{A}$ and $\mathcal{H}_{B}$ are finite dimensional Hilbert spaces. Then, using the Schmidt decomposition theorem, $\ket{\Psi}_{AB}$ can be written in the form
\begin{equation}
    \ket{\Psi}_{AB}=\sum_{j=0}^{d-1}\sqrt{\lambda_{j}}\ket{j}_{A}\ket{j}_{B},
\end{equation}
where $\{\ket{j}_{A}\}$, $\{\ket{j}_{B}\}$ are orthonormal vectors corresponding to subsystems $A$ and $B$, respectively and $\{\lambda_{n}\}$ are the Schmidt coefficients (non-negative real numbers) with $\sum_{n}\lambda_{n}=1$, and $d\leq{\mathrm {min}}\{{{\mathrm{ dim}}\left({\cal{H}}_{A}\right)},{{\mathrm{ dim}}\left({\cal{H}}_{B}\right)}\}$. 
For mixed state, the definition of entanglement is more involved.
Consider a density operator $\rho_{AB}$  which is a positive, Hermitian operator
with unit trace acting  on ${\cal H} = {\cal H}_A  \otimes {\cal H}_B$. The state $\rho_{AB}$ is called separable if
there exists an ensemble $\{ p_i,  \rho_{iA} \otimes \rho_{iB} \}$ with 
$$ \rho_{AB} = \sum_i p_i \rho_{iA} \otimes \rho_{iB}. $$
If $\rho_{AB}$ cannot be expressed as a separable form then it is entangled.

\section{Measuring entanglement of pure state from single copy}\label{sub4}
 
 In this section, we propose two novel methods to measure the linear entanglement entropy and the negativity by choosing suitable unitary operators with single copy at the input port of
 the interferometric set-up.
 
 \subsection{Measuring Linear Entanglement Entropy with Single Copy}
 To measure the purity of a state, usually one has to 
estimate the quantum state and from the classical description of the density operator one can 
 extract the purity by classical evaluations. However, it should be noted that the estimation
of purity does not require the knowledge of full density operator. Therefore, prior state estimation procedure followed by classical calculations, in general, are inefficient. To overcome this, it was proposed that direct measurement of purity can be carried out with the help of quantum networks which is essentially an interferometric scheme \cite{Horodecki2002}. However, this proposal needs two copies of the input state $\rho$ to measure the purity. If one has a limited supply of quantum systems, it is desirable  to have more efficient scheme which can reduce the number of copies at the input port of the interferometer. 

In this section, we achieve this precisely where having access to one copy of the density operator $\rho$, we can measure the purity $\tr(\rho^2)$, and the linear entanglement entropy of any pure bipartite quantum state. Given a pure bipartite state $\ket{\Psi}_{AB}$ in Hilbert space $\mathcal{H}_{A}\otimes\mathcal{H}_{B}$, the entanglement content can be quantified using the linear entropy \cite{U00} which is given by
 \begin{equation}
 \label{concurrence}
 \mathcal{E}(\ket{\Psi}) = \left(1-\tr(\rho_{A}^{2})\right),
 \end{equation}
 where $\rho_{A}$ is the reduced density matrix of the subsystem $A$. 
 The linear entropy is a valid measure of entanglement similar to the von Neumann entropy of entanglement. This is also concave and unitarily invariant. 
 If one can measure $\tr(\rho_{A}^{2})$ in an experiment, then it is possible to measure the linear entropy. For $d=2$, the linear entropy is the same as the concurrence of the bipartite quantum state. 
 Here, we propose a direct method of measuring the linear entropy with single copy of the input state $\rho_A$ along with a single unitary operation. This method works even if we do not know the pure bipartite entangled state.
 
To achieve this, we need a unitary operator where the additional power comes from the ability to verify a state instead of knowing the state completely. Similar argument has been used in the quantum state restoration method using single-copy tomography \cite{Farhi2010}. For a single copy of an unknown quantum state $\ket{\Psi}_{AB}$ a black-box (oracle) can exist which can implement the unitary $U=(\mathbb{I}-2P)$, where $P = \ket{\Psi}_{AB}\bra{\Psi}$. Since, we are not interested in knowing the state, the no-cloning theorem \cite{Wooters1982} does not apply here.  Using this oracle operator $U$, we will show how one can measure the linear entropy  $ \left(1 -\tr(\rho^2_{A})\right)$ from the interference visibility. Suppose, we want to measure the entanglement content of a pure bipartite state $\ket{\Psi}_{AB}$. Instead of sending the full state, we send one of the local subsystem $\rho_A$ and another system which is prepared in a maximally mixed state $\frac{\mathbb{I}_{B}}{d}$.
Let $U = (\mathbb{I}-2P)$ acts on one arm of the interferometer where we have fed the state $\rho_{AB} = \rho_{A}\otimes \frac{\mathbb{I}_{B}}{d}$. This unitary operation and the state $\rho_{AB}$ will be used in Eq.~(\ref{intensity}) to determine $\tr(U\rho)$ from the interference pattern as given by Eq.~(\ref{intensity}).  The protocol to measure the linear entanglement entropy described above can be depicted in the following diagram.

\begin{framed}
\begin{center}
\begin{tikzpicture}
\node at (-0.5,0) {$\ket{0}$}; 
\draw (0,0) -- (2,0);
\draw (2,-0.5) -- (2,0.5) -- (3,0.5) -- (3,-0.5) -- (2,-0.5);
\node at (2.5,0) {$H$};
\draw (3,0)--(4,0);
\node at (4,0.5) {$\phi$};
\filldraw[black] (4,0) circle (2pt);
\draw (4,0)--(5,0);
\node at (5.5,0) {$H$};
\draw (6,0) -- (6.65,0);
\draw[black](7,0) circle (10pt);
\node at (7,0) {$M$};
\draw (5,-0.5) -- (5,0.5) -- (6,0.5) -- (6,-0.5) -- (5,-0.5);
\draw (4.5,0)--(4.5,-1.5);
\node at (4.5,-2) {$\scalemath{0.8}{I-2P}$};
\draw (4,-1.5) -- (5,-1.5) -- (5,-2.5) -- (4,-2.5) -- (4,-1.5);
\draw (0,-1.75) -- (4,-1.75);
\draw (0,-2.15) -- (4,-2.15);
\draw (5,-2) -- (6,-2);
\node at (-0.5,-1.75) {$\rho_{A}$};
\node at (-0.5,-2.15) {$\mathbb{I}_{B}$};
\end{tikzpicture}
\label{figure}
\end{center}
Here we depict the interferometric setup for measuring entanglement. The entanglement of state $\ket{\Psi}_{AB}$ is measured through the visibility of the interference
pattern according to Eq.~\eqref{visibilityconcurrence11}. For any two-qubit pure state the visibility is a direct measure of linear entanglement entropy.
\end{framed}

For this choice of the oracle unitary operator and the initial state, the interferometric visibility is modified as
 \begin{align}
  V & = \left|\tr(U\rho_{A}\otimes \frac{\mathbb{I}_{B}}{d})\right|\nonumber\\
    & = \left|\tr\left(\left(\mathbb{I}-2P\right)\left(\rho_{A}\otimes \frac{\mathbb{I}_{B}}{d}\right)\right)\right|\nonumber\\
    \label{visibilityconcurrence}
    & = 1-\frac{2}{d} \tr(\rho^2_{A}).
    \end{align}
Using the interference visibility we can measure the linear entanglement entropy as given by
\begin{equation}
    \mathcal{E}(\ket{\Psi}) = 1 - \frac{d}{2}(1-V).\label{visibilityconcurrence11}
\end{equation}   
Our method works even for unknown pure entangled state. As explained, the oracle has the capability of implementing the unitary $U=(\mathbb{I}-2P)$ even for an unknown state $P=\ket{\Psi}_{AB}\bra{\Psi}$. We only need the quantum state corresponding to subsystem $A$, at the input port of the interferometer along with an additional maximally mixed state. This shows that 
for any arbitrary two qubit pure entangled state, we have $\mathcal{E}(\ket{\Psi}) = V$. Thus, for any pure bipartite two qubit state, the the interference visibility is a direct measure of entanglement.
It may be noted that, in general, to evaluate the spectrum of any density matrix we need to estimate $(d^2 - 1 )$ parameters. However, our method can estimate the linear entropy directly without full quantum 
state tomography. Another important feature of this method is that we need single copy of the input state $\rho$ at the input port of the interferometer, and therefore it is also resource economical.
Our proposal can be designed as a portable device to measure the entanglement of any pure bipartite state from the visibility, thus paving towards the notion of an entanglement meter.

Now, we extend our analysis for the case of mixed states. Consider a bipartite system $AB$ in mixed state. Then, the density operator $\rho_{AB}$ can be written as convex mixture of pure states $\Psi_{i}$ with mixing probabilities $p_{i}$, i.e., $\rho_{AB}  =\sum_{i}p_{i}\Psi_{iAB}$, where each $\Psi_{iAB}$ is one dimensional projector, i.e., $\Psi_{i} = \ket{\Psi_i}_{AB}\bra{\Psi_i}_{AB}$, and $\sum_{i}p_{i} =1$. Using the convex roof construction, the linear entropy can also be used as a measure of entanglement for mixed states~\cite{Toth2015}. 
\begin{definition}
For a bipartite mixed state $\rho_{AB}$ with decomposition $\{p_{i},\ket{\Psi_{i}}\}$, the convex roof extended linear entropy is defined as~\cite{Toth2015}
\begin{equation}
    \mathcal{E}_{c}(\rho_{AB}) =\min_{\{p_{i},\ket{\Psi_{i}}\}}\left(p_{i}\mathcal{E}(\ket{\Psi_{i}})\right),
\end{equation}
here minimum is taken over all decomposition of $\rho_{AB}$,  $\mathcal{E}(\ket{\Psi_{i}}) = \left(1-\tr(\rho^2_{Ai})\right)$ is the linear entropy of pure state $\ket{\Psi_{i}}$, and $\rho_{Ai} = \tr_{B}(\ketbra{\Psi_{i}}{\Psi_{i}})$.
\end{definition} 
The reduced density operator $\rho_{A}$ is given by $ \rho_A = \tr_{B}(\sum_{i}p_{i}\ketbra{\Psi_{i}}_{AB}{\Psi_{i}})$. From the definition of convex roof extended linear entropy, we have the following inequality
\begin{align}
    \mathcal{E}_{c}(\rho_{AB})\leq \sum_{i}p_{i}(1-\tr(\rho^2_{Ai})).
\end{align}
Now, we show that the term on the right hand side of the above inequality is upper bounded by $1-\tr(\rho^2_{A})$.  Using the Cauchy-Schawrz inequality and the fact that geometric mean is less than or equal to arithmatic mean we obtain 
\begin{equation}
\label{inequality}
\tr(AB)\leq \sqrt{\tr(A^2)\tr(B^2)}\leq \frac{1}{2}(\tr(A^{2})+\tr(B^{2})),
\end{equation}
 where $A$ and $B$ are Hermitian operators.  Using Eq.~(\ref{inequality}) we obtain
\begin{align}
\label{stateinequality}
\tr(\rho^2_{A}) & \leq \frac{1}{2}\sum_{i,j}p_{i}p_{j}\left(\tr(\rho^2_{Ai})+\tr(\rho^2_{Aj})\right).
\end{align}
Summing over $i(j)$ in the first(second) term of the R.H.S of Eq.~(\ref{stateinequality}), we obtain
\begin{align}
    \tr(\rho^2_{A}) & \leq \sum_{i}p_{i}\tr(\rho^2_{Ai}).
    \end{align}
    This shows that the convex the roof extended linear entropy is upper bounded by the following
\begin{align}
\label{desired_inequality}
   \mathcal{E}_{c}(\rho_{AB}) &\leq  1-\tr(\rho^2_{A}).
\end{align}
We have already provided a method to measure $\tr(\rho^2_{A})$ from interferometric visibility. Using Eq.~\eqref{visibilityconcurrence} and inequality~\eqref{desired_inequality}, we obtain the following inequality for $ \mathcal{E}_{c}(\rho_{AB})$ in terms of interferometric visibility
\begin{align}
    \mathcal{E}_{c}(\rho_{AB})\leq 1-\frac{d(1-V)}{2}.
\end{align}
From the above inequality we then obtain an upper bound on convex roof extended linear entropy in terms of interferometric visibility $V$ and $d$. This shows that, for an arbitrary two-qubit mixed state $\rho_{AB}$, we have $\mathcal{E}_{c}(\rho_{AB})\leq V$, i.e., the visibility gives an upper bound to the convex roof extended linear entropy for $\rho_{AB}$.

The above procedure is one of the examples of how we can determine entanglement of pure quantum state by having access to one copy of the any one of the reduced density operator. We do not need the full knowledge of the joint state. This shows the true power of quantum interference in determining the entanglement content. In addition, our method is more efficient in resource consumption compared to Refs.~\cite{Horodecki2002,Bovino2005,Mintert2007,EA+02} as mentioned previously.

\subsection{Measuring Negativity with Single Copy}

Note that the above protocol depends on an oracle which produces a unitary $ U = (\mathbb{I}-2P)$ given a density matrix of a pure state $P$. In what follows, we provide another method where, given a bipartite state $\rho_{AB}$, and a unitary operation $U={\mathrm e}^{i\theta X_{A}}\otimes { \mathrm e}^{i\theta X_{B}}$ such that $\tr(U\rho)$ contains the signature of entanglement in terms of negativity measure of entanglement. Since the linear entropy and the negativity only differ by a multiplicative factor for any pure state, one can implement either the first or the second protocol depending upon the resource one has at the disposal. However, in case of mixed states, we will show that there exists class of mixed state, for which, the second method provides the exact value of the negativity whereas the first method only provides a lower bound.

The negativity is an entanglement monotone which is derived from the positive partial transpose (PPT) criterion for the separability of a bipartite
quantum states. For a pure bipartite state $\ket{\Psi}= \sum_{j=0}^{d-1}\sqrt{\lambda_{j}}\ket{j}_{A}\ket{j}_{B}$, its negativity is given by~\cite{VW02} 
\begin{equation}
\label{negativity}
    \mathcal{N}(\ket{\Psi}) = \frac{1}{2}\sum_{i\neq j} \sqrt{\lambda_{i} \lambda_{j}},\end{equation}
where $\lambda_{i}$'s are Schmidt coefficients.
Now, consider an operator 
$$X=\sum_{i,j=0;i\neq j}^{d-1}\ketbra{i}{j}=\sum_{i<j}\underbrace{\left(\ketbra{i}{j}+\ketbra{j}{i}\right)}_{G_{ij}}.$$  This operator is Hermitian in any finite dimension. In addition, for $d=2$, this operator is also unitary as $X^2 =\mathbb{I}$.
First, we describe how to measure the negativity of any pure bipartite two qubit state directly using the interference visibility. For example, consider the input state as an arbitrary pure two qubit state $\ket{\Psi}_{AB}$ in the Mach-Zehnder interferometer and apply the unitary operator $ U= X_A \otimes X_B$ in one arm of the interferometer.
The visibility is given by
\begin{align}
\label{negativityvisibilitytwoqubit}
|\tr(U \rho_{AB})| = \bra{\Psi}X_{A}\otimes X_{B}\ket{\Psi} & = 2\mathcal{N}(\ket{\Psi}).
\end{align}
Thus, by looking at the interference visibility we can infer the negativity and hence the entanglement content of any pure bipartite two-qubit state. Note that Eq.~(\ref{negativityvisibilitytwoqubit}) is only valid for $d=2$. We will show how the visibility is modified for $d>2$.\\

The observable $X$ has the following property for any positive integer $n>2$, i.e., 
\begin{equation}
\label{observableproperty}
X^{n}=(d-1)X^{n-2}+(d-2)X^{n-1}.
\end{equation}
To prove Eq.~(\ref{observableproperty}) we only need to prove that $X^{2}$ can be written as linear combination of $X$ and identity operator $\mathbb{I}$ as 
\begin{equation}
\label{obsproperty}
    X^{2}=(d-1)\mathbb{I}+(d-2)X.
\end{equation}
From Eq.~(\ref{obsproperty}), we can also see that for $d=2$, $X^{2}=\mathbb{I}$. Note that
\begin{equation}
\label{obsprop}
X^{2}=\sum_{i,j,l,m=0;i\neq j,l\neq k}^{d-1}\ketbra{i}{l}\delta_{jm}.
\end{equation}
Now, if we take $i=l$ in the above equation, then it can be easily checked that there are $(d-1)$ values of $j$ and $m$, for which $\delta_{jm}=1$. Similarly, if we take $i\neq l$, then there are $(d-2)$ values of $j$ and $m$ for which $\delta_{jm}=1$. Therefore, Eq.~(\ref{obsprop}) can be rewritten as
\begin{align}
X^{2}& = (d-1)\sum_{i,l=0;i=l}^{d-1}\ketbra{i}{l}+(d-2)\sum_{i,l=0;i\neq l}^{d-1}\ketbra{i}{l} \nonumber\\
\label{obsprob1}
& = (d-1)\mathbb{I}+(d-2)X.
\end{align}
If we multiply $X^{n-2}$ on both the sides of the Eq.~(\ref{obsprob1}), then we obtain Eq.~(\ref{observableproperty}). Note that Eq.~(\ref{observableproperty}) is valid for $n>2$. We can extend it to all non-zero positive integer powers of $n$ by considering the following generalization
\begin{equation}
\label{recursivefinal}
X^{n}=f_{n}X+g_{n}\mathbb{I}.
\end{equation}
Here, $f_{n}$ and $g_{n}$ are real-valued functions which depend on the dimension of the Hilbert space $\mathrm{dim}(\cal{H})$ and the power of $X$, $n$. Now, considering $n=2$ in Eq.~(\ref{recursivefinal}) and comparing it with Eq.~(\ref{obsprob1}), we obtain $f_{2}=(d-2)$ and $g_{2}=(d-1)$. Again, considering $n=1$, we then obtain $f_{1}=1$ and $g_{1}=0$.  If we consider Eq.~(\ref{recursivefinal}) for $(n-1)$th power, i.e., 
$$X^{n-1}=f_{n-1}X+g_{n-1}\mathbb{I}$$
and multiply it with $X$ on both the sides, then using Eq.~(\ref{obsprob1}) on the R.H.S, we obtain the following relations
\begin{subequations}
\begin{align}
\label{grecursion}
g_{n}&=(d-1)f_{n-1},\\
\label{frecursion}
f_{n}&=(d-2)f_{n-1}+(d-1)f_{n-2}.
\end{align}
\end{subequations}
Note that $f_{n}$ follows a general Fibonacci-like recursive relation. This sequence is known as the Lucas sequence of first kind and its solution is given as follows \cite{proofthatcounts}
\begin{equation}
\label{solfn}
f_{n}=\frac{(d-1)^{n}-(-1)^{n}}{d}.
\end{equation}
Similarly, we obtain
\begin{equation}
\label{solgn}
g_{n}=\frac{d-1}{d}\left[(d-1)^{n-1}-(-1)^{n-1}\right].
\end{equation}

 Now defining 
\begin{align}
X_{A}&=\sum_{i<j}G^{A}_{ij}\label{observableA},\\
X_{B}&=\sum_{i<j}G^{B}_{ij}\label{observableB},
\end{align}
 we obtain
\begin{align}
\label{jointexpectationvalue}
\bra{\Psi}X_{A}\otimes X_{B}\ket{\Psi} & = 2\mathcal{N}(\ket{\Psi}) \\
\label{marginalexpectationvalue}
\bra{\Psi}X_{A}\otimes \mathbb{I}_{B}\ket{\Psi} & = \bra{\Psi}\mathbb{I}_{A}\otimes X_{B}\ket{\Psi} = 0.
\end{align}

Now, consider an unitary $U={\mathrm e}^{i\theta X}$. Expanding the exponential and using Eqs.~(\ref{recursivefinal}), (\ref{solfn}) and (\ref{solgn}) we obtain
\begin{align}
\label{unitarynegativity}
{ \mathrm e}^{i\theta{X}} & = \sum_{n=0}^{\infty}\frac{i^{n}\theta^{n}X^{n}}{n!}\\
& = \scalemath{0.75}{\mathbb{I}+\frac{1}{d}\sum_{i=1}^{n}\frac{i^{n}\theta^{n}}{n!}\bigg[(d-1)^{n}\left(X+\mathbb{I}\right)+(-1)^{n}\left((d-1)\mathbb{I}-X\right)\bigg]}\\
& = \scalemath{0.75}{\frac{1}{d}\sum_{i=0}^{n}\frac{i^{n}\theta^{n}}{n!}\bigg[(d-1)^{n}\left(X+\mathbb{I}\right)+(-1)^{n}\left((d-1)\mathbb{I}-X\right)\bigg]}\\
& = \frac{1}{d}\bigg[{\mathrm e}^{i\theta(d-1)}(X+\mathbb{I})+{\mathrm e}^{-i\theta}((d-1)\mathbb{I}-X)\bigg]\\
\label{negativityunitary}
& = \frac{{ \mathrm e}^{-i\theta}}{d}\bigg[\left({\mathrm e}^{i\theta d}-1\right)X+\left({\mathrm e}^{i\theta d}+(d-1)\right)\mathbb{I}\bigg],
\end{align}
where we use 
$$\mathbb{I}=\frac{1}{d}[(d-1)^{0}\left(X+\mathbb{I}\right)+(-1)^{0}\left((d-1)\mathbb{I}-X\right)].$$

Given a pure state $\rho_{AB}=\ketbra{\Psi}{\Psi}$ and $\ket{\Psi}_{AB} =\sum_{j}\sqrt{\lambda_{j}}\ket{j}_{A}\ket{j}_{B}$, the quantity $\tr(U_{A}\otimes U_{B}\rho_{AB})$ contains the following three terms $\bra{\Psi}X_{A}\otimes X_{B}\ket{\Psi}$, $\bra{\Psi}X_{A}\otimes \mathbb{I}_{B}\ket{\Psi} = \bra{\Psi}\mathbb{I}_{A}\otimes X_{B}\ket{\Psi}$. Eqs.~(\ref{marginalexpectationvalue}) and (\ref{jointexpectationvalue}) shows that the first term is proportional to negativity, $\mathcal{N}(\ket{\Psi})$, second and third terms are zero. We then have the following
\begin{align}
\label{unitaryexpectation}
\tr(U\rho) & = \frac{{\mathrm e}^{-2i\theta}}{d^{2}}\bigg[2\left({\mathrm e}^{i\theta d}-1\right)^{2}\mathcal{N}+\left( {\mathrm e}^{i\theta d}+(d-1)\right)^{2}\bigg],
\end{align}
where $U=U_{A}\otimes U_{B}$, $U_{A}$ and $U_{B}$ are given by Eq.~(\ref{negativityunitary}).

If we take $\theta=\frac{\pi}{d}$ in the above equation then we have
\begin{align}
\label{unitaryexpectationforspectheta}
\tr(U\rho) & = \frac{{\mathrm e}^{-2i\pi/d}}{d^{2}}\bigg[8\mathcal{N}+(d-2)^{2}\bigg].
\end{align}
The visibility is then given as
\begin{equation}
\label{visibilitynegativity}
|\tr(U\rho)| = \frac{1}{d^{2}}\left(8\mathcal{N}+(d-2)^{2}\right).
\end{equation}
Note that for $d=2$ we obtain Eq.~(\ref{negativityvisibilitytwoqubit}) from Eq.~(\ref{visibilitynegativity}). This shows us that even for higher dimensional pure bipartite state, we can measure the negativity directly using the interference visibility.\\

The above protocol is not restricted to pure state. To see that, consider the following mixed state which is a mixture of maximally entangled state and a classical-classical state
\begin{equation}
\label{cna}
\rho_{AB}=x\ketbra{\Phi^{+}}{\Phi^{+}}+\frac{(1-x)}{d}\sum_{j=0}^{d-1}\ketbra{j}{j}\otimes\ketbra{j}{j},
\end{equation}
where $x$ is a real number and $x\in[0,1]$. Here, $\ket{\Phi^{+}}=\frac{1}{\sqrt{d}}\sum_{j}\ket{j}_{A}\ket{j}_{B}$ is the maximally entangled state of two qudit. The entanglement of the above state is given by \cite{hm+16} 
\begin{equation}
\mathcal{N}(\rho_{AB}) =\frac{x(d-1)}{2}.
\end{equation}
Now, it can be shown that if $X_{A}$ and $X_{B}$ are given by Eqs.~(\ref{observableA}) and (\ref{observableB}), then the joint observable $X_{A}\otimes X_{B}$, satisfies the relationships given by Eqs.~(\ref{jointexpectationvalue}) and (\ref{marginalexpectationvalue}). In particular,
\begin{align}
\label{jointexpectationvaluemixed}
\tr(X_{A}\otimes X_{B}\rho_{AB}) & = \frac{2\mathcal{N}(\rho_{AB})}{d-1}\\
\label{marginalexpectationvaluemixed}
\tr(X_{A}\otimes \mathbb{I}_{B}\rho_{AB}) & = \tr(\mathbb{I}_{A}\otimes X_{B}\rho_{AB}) = 0.
\end{align}
Therefore, the entanglement of the state given by Eq.~(\ref{cna}) can be determined from the interferometric setup.\\

The results presented here can pave the way towards design of an entanglement meter where we can have a portable device inside which we have the Mach-Zehnder interferometer with suitable unitary operators in place. By sending single copy of the local system or joint system at the input port of the entanglement meter, we can directly measure the entanglement between the pairs of subsystems. This can have several industrial applications in quantum technology. For example, if a private company is supplying maximally entangled pairs, then to convince a buyer, company can give one of the pair to the user  along with the entanglement meter which has the sequence of unitaries along with the oracle. Using the entanglement meter, the end user can verify that indeed the supplied pair is maximally entangled state.
We have described two protocols for determining the amount of entanglement. The first protocol requires a single copy of the reduced state and the second protocol uses a single copy of the joint state. In the case of pure states, one can implement either the first or the second protocol. However, in the case of mixed states, the first protocol yields an upper bound on the convex roof extended entanglement measure. On the other hand, the second protocol yields the exact value of the entanglement measure for some class of mixed states.

\begin{figure}[h!]
    \centering
     \includegraphics[width=7cm]{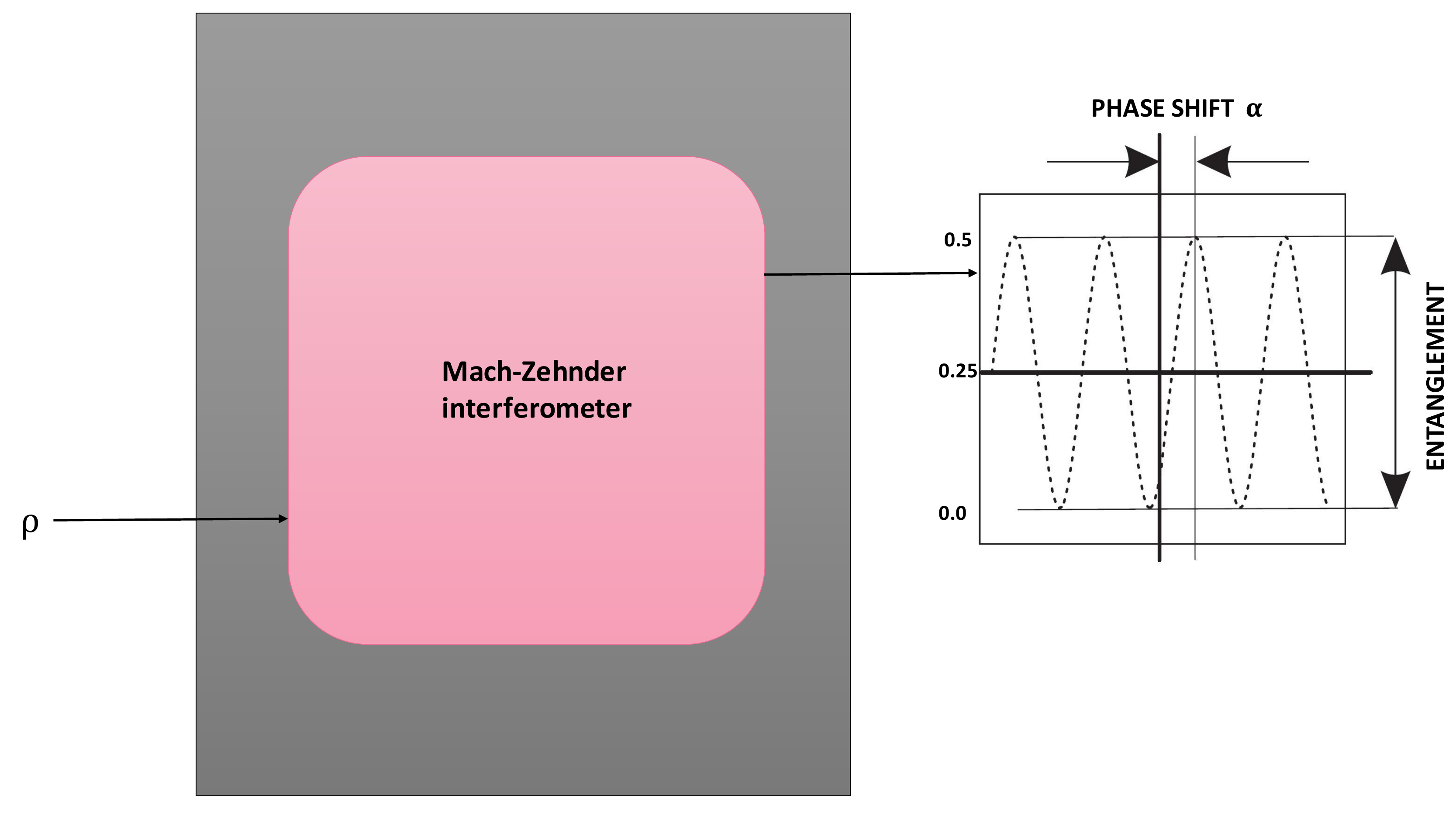}
    \caption{Schematic of the entanglement meter. By looking at the interference pattern we can say how much entanglement is present in 
    the two-qubit pure state.}
  \label{fig:Entanglement_meter}
\end{figure}
\section{Entanglement Detection from Interferometric set-up}
Finding out whether a given state is entangled or separable is one of the fundamental and crucial problems in the theory of entanglement. For qubit-qubit and qubit-qutrit systems there exist a necessary and sufficient condition
for separability based on PPT criteria~\cite{Horodecki1996,Peres1996,Horodecki2009,Sreetama2017}. For multipartite systems or higher dimensional systems the most general method to detect entangled states is via entanglement witness~\cite{CG14}. An alternative approach to detect entanglement is via measuring suitably defined statistical correlations (for example, mutual predictability \cite{SH+12} or Pearson correlation \cite{Maccone2015}) in complementary observables.  The notion of ``mutual predictability"  is a complementary correlation which can be used to detect entanglement in multipartite systems or higher dimensional systems as well as for continuous variable systems~\cite{SH+12}. Another approach to detect entanglement is to use a modification of the Bell scenario \cite{bell} which demonstrate that all entangled state violates a inequality involving joint probabilities \cite{Buscemi2012}. In this section our objective is to demonstrate how some of the existing entanglement detection methods can be implemented in a interferometric set-up. In particular, we have chosen the entanglement detection process involving mutual predictability \cite{SH+12}. Instead of performing local measurements in complementary bases we have constructed a suitable unitary which yields the value of mutual predictability directly. Furthermore, we will also discuss how some class of entanglement witness operators can be implemented in a interferometric setup.

\subsection{Determination of Mutual Predictability from
Interferometric set-up}\label{subA}
In Ref.~\cite{gr+13,SH+12}, it was shown that there is intimate connection between the mutually unbiased bases~(MUBs) and detection of entanglement for composite systems in high-dimensional Hilbert space. In particular by performing local measurements in mutually unbiased bases one can detect whether a state is entangled or not by using a statistical quantity called the mutual predictability. In addition, 
the approach given in Ref.~\cite{SH+12} can provide necessary and sufficient criteria for separability if a complete set of MUBs is available for the local subsystems. In this section, we show that without doing explicit local measurements of complete set of MUBs, we can construct a unitary operation which yields the value of mutual predictability in terms of the interferometric visibility and hence can detect entanglement.

 Consider a bipartite state $\ket{\Psi}_{AB}\in \mathcal{H}_{A}\otimes\mathcal{H}_{B} $ and observables ${\cal{O}}_{A}$ and ${\cal{O}}_{B}$ for each subsystem. The eigenvectors of ${\cal{O}}_{A}$ and ${\cal{O}}_{B}$ can be taken as $\{\ket{k}_{A}\}$ and $\{\ket{l}_{B}\}$, where $  l,k \in \{0,1,2,3,.......d-1\} $ and $d\leq{\mathrm {min}}\{{{\mathrm{ dim}}\left({\cal{H}}_{A}\right)},{{\mathrm{ dim}}\left({\cal{H}}_{B}\right)}\}$.
We can define mutual predicatibility as 
\begin{equation}
\label{mutualpredict}
 C_{{\cal{O}}_{A},{\cal{O}}_{B}} = \sum_{k} {_A}\bra{k} {_{B}}\bra{k} \rho_{AB} \ket{k}_{A} \ket{k}_{B}.
\end{equation}
Now, if we consider $m$ mutually unbiased dobservables for each subsystem as 
$\{{\cal{O}}_{A_{1}},{\cal{O}}_{A_{2}},......{\cal{O}}_{A_{m}}\}$ and $\{{\cal{O}}_{B_{1}},{\cal{O}}_{B_{2}},......{\cal{O}}_{B_{m}}\}$, then for separable state we have \cite{SH+12}
\begin{equation}
\label{mutualpredictability}
\sum_{i=1}^{m}C_{{\cal{O}}_{A_{i}},{\cal{O}}_{B_{i}}} \leq 1 + \frac{m-1}{d}.
\end{equation}
If any state violates the upper bound given in Eq.~(\ref{mutualpredictability}) then it is entangled. If $m=d+1$, then the above criteria a necessary and sufficient to detect separability of any bipartite state \cite{SH+12} . In what follows, we will provide a scheme to obtain $C_{{\cal{O}}_{A_{i}},{\cal{O}}_{B_{i}}} $ from the interference experiment.
Note that given any projector $\Pi$, the oracle can be represented by a unitary operator
$$U = \mathbb{I}-2\Pi.$$
Now let us consider two observable ${\cal{O}}_{A}$ and ${\cal{O}}_{B}$ with the  basis vectors are  $\{\ket{k_{A}}\}$ and $\{\ket{l_{B}}\}$.
Now, Let us define $U_{k}=\mathbb{I}\otimes\mathbb{I}-2\ket{k}_{A}\bra{k}\otimes\ket{k}_{B}\bra{k}$. We can then consider the following unitary operator 
\begin{align}
 U_{{\cal{O}}_{A},{\cal{O}}_{B}} & = \prod_{k}U_{k}\nonumber\\
 & = \prod_{k} \left(\mathbb{I}\otimes\mathbb{I}-2\ket{k}_{A}\bra{k}\otimes\ket{k}_{B}\bra{k}\right)\nonumber\\
 & = \left(\mathbb{I} \otimes \mathbb{I} - 2\sum_{k} \ket{k}_{A}\bra{k}\otimes\ket{k}_{B}\bra{k}\right).\label{equ:unitary_for_mutually_unbiased_observables}
 \end{align}
 Consider the state $\rho_{AB}$ at the input port of the interferometer. Let $U_{{\cal{O}}_{A},{\cal{O}}_{B}}$ act on one arm of the interferometer.
 Now we calculate $\tr(\rho_{AB} U_{{\cal{O}}_{A},{\cal{O}}_{B}})$ which is given by
 \begin{align*}
 \tr\left(\rho_{AB} U_{{\cal{O}}_{A},{\cal{O}}_{B}}\right) &= \scalemath{.9}{\tr\left(\Big(\mathbb{I} \otimes \mathbb{I} - 2\sum_{k} \ket{k}_{A}\bra{k}\otimes\ket{k}_{B}\bra{k}\Big)\rho_{AB}\right)}\\
 & = 1 -  2\sum_{k}\tr \left( \left(\ket{k}_{A}\bra{k}\otimes\ket{k}_{B}\bra{k}\right)\rho_{AB}\right)\\
 & = 1 - 2\sum_{k} {_A}\bra{k} {_{B}}\bra{k} \rho_{AB} \ket{k}_{A} \ket{k}_{B}\\
 & = 1- 2C_{{\cal{O}}_{A},{\cal{O}}_{B}}.
 \end{align*}
 The above equation shows that the mutual predictability can be obtained without doing measurement on the system itself and reveals another power of interferometer setup.
 Since $0 \leq C_{{\cal{O}}_{A},{\cal{O}}_{B}}\leq 1$, we have
 \begin{align*}
 \left|\tr(\rho_{AB} U_{{\cal{O}}_{A},{\cal{O}}_{B}})\right|  = |1 - 2C_{{\cal{O}}_{A},{\cal{O}}_{B}}|.
  \end{align*}
  If $C_{{\cal{O}}_{A},{\cal{O}}_{B}}>\frac{1}{2}$, then the visibility will be $(2C_{{\cal{O}}_{A},{\cal{O}}_{B}}-1)$, if it is less than half then visibility will be $(1-2C_{{\cal{O}}_{A},{\cal{O}}_{B}})$. Furthermore, the phase shift, $\alpha=\arg[\tr(U_{\mathcal{O}_{A},\mathcal{O}_{B}}\rho_{AB})]$ will be zero in this case. We can rewrite above equation as 
 \begin{align}
 C_{{\cal{O}}_{A},{\cal{O}}_{B}} = \frac{1}{2}\left(1 \pm \left|\tr(\rho_{AB} U_{{\cal{O}}_{A},{\cal{O}}_{B}})\right|\right).\label{equ:visibility_as_mutual_predactibility}
  \end{align} 
Similarly, if we consider $m$ mutually unbiased observables for each subsystem as 
$\{{\cal{O}}_{A_{1}},{\cal{O}}_{A_{2}},......{\cal{O}}_{A_{m}}\}$ and $\{{\cal{O}}_{B_{1}},{\cal{O}}_{B_{2}},......{\cal{O}}_{B_{m}}\}$, then for each pair of observables $\{{\cal{O}}_{A_{i}},{\cal{O}}_{B_{i}}\}$ we have a unitary $U_{{\cal{O}}_{A_i},{\cal{O}}_{B_i}}$ similar to Eq.~\eqref{equ:unitary_for_mutually_unbiased_observables} and we can obtain mutual predictability in terms of $\left|\tr(\rho_{AB} U_{{\cal{O}}_{A_i},{\cal{O}}_{B_i}})\right|$ as 
 \begin{align}
 C_{{\cal{O}}_{A_i},{\cal{O}}_{B_i}} = \frac{1}{2}\left(1 \pm \left|\tr(\rho_{AB} U_{{\cal{O}}_{A_i},{\cal{O}}_{B_i}})\right|\right),
  \end{align} 
like Eq.~\eqref{equ:visibility_as_mutual_predactibility}. Taking summation over $i$ from 1 to $m$ on both the sides of the above equation, we then obtain
 \begin{align}
 \label{visibilitypredictability}
  \sum_{i}^{m} C_{{\cal{O}}_{A_i},{\cal{O}}_{B_i}} = \frac{1}{2}\left(m\pm\sum_{i}^{m}\left|\tr(\rho_{AB} U_{{\cal{O}}_{A_i},{\cal{O}}_{B_i}})\right|\right).
  \end{align} 
Since $\left|\tr(\rho_{AB} U_{{\cal{O}}_{A_i},{\cal{O}}_{B_i}})\right|$ is real  and positive, it implies that $\arg[\tr(U_{\mathcal{O}_{A},\mathcal{O}_{B}}\rho_{AB})]$ will vanish and phase will be constant for all unitary operators  $\{U_{\mathcal{O}_{A_i},\mathcal{O}_{B_i}}\}$. Let us denote $\left|\tr(\rho_{AB} U_{{\cal{O}}_{A_i},{\cal{O}}_{B_i}})\right|=V_{i}$. Now, using inequality \eqref{mutualpredictability}, Eq.~\eqref{intensity} and the above equation, we obtain the separability criteria in terms of $\left|\tr(\rho_{AB} U_{{\cal{O}}_{A_i},{\cal{O}}_{B_i}})\right|$
 \begin{equation}
 \label{visibilityinequality}
    2\left(1+\frac{(m-1)}{d}\right)-m \geq \sum_{i=1}^{m}V_{i} \geq m-2\left(1+\frac{(m-1)}{d}\right).
    \end{equation}
 Here $V_{i}$ is the visibility obtained after implementation of $i$th unitary. If a state $\rho_{AB}$ violates the above inequality then the state is entangled. Let us study the above inequality in some detail. Note that $\sum_{i}V_{i}$ is always positive. However, it may happen that either the lower or upper bound can be negative. In that case, only one part of the inequality is valid for a separable state.\\
 
 To give an example, consider the case where there are  $d+1$ mutually unbiased bases, i.e., $m=d+1$. In that case, the mutual predictability is a necessary and sufficient condition for separability \cite{SH+12} and Eq.~(\ref{visibilityinequality}) can be modified as
 \begin{equation*}
(d-3)\leq \sum_{i=1}^{d+1}V_{i} \leq (3-d).
 \end{equation*}
 We can see that the upper bound is less than zero for $d>3$, thus it is unattainable by $\sum_{i}^{d+1}V_{i}$ and the entanglement detection can be done through the lower bound on $\sum_{i=1}^{d+1}V_{i}$ i.e.,
 \begin{equation}
\label{completevisibilityinequality}
\sum_{i=1}^{d+1}V_{i}\geq (d-3).
 \end{equation}
 If the quantum state violates the above inequality for $(d+1)$ mutually unbiased observables then it is entangled. Similarly, in the case of $m=2$, the separable states satisfy 
 \begin{equation}
\label{twovisibility}
V_{1}+V_{2}\geq \frac{1}{d}.
 \end{equation}
  As an example of how Eq.~(\ref{visibilityinequality}) can detect entanglement consider the $d$ dimensional isotropic state given by $\rho_{I}=x\ketbra{\Phi^{+}}{\Phi^{+}}+\frac{(1-x)}{d^{2}}\mathbb{I}\otimes\mathbb{I}$, where $\ket{\Phi^{+}}=\frac{1}{\sqrt{d}}\sum\ket{j}_{A}\ket{j}_{B}$ and $x$ is a real number between zero to one. The isotropic state is separable if $x\leq 1/(d+1)$. Note that isotropic state is $U\otimes U^{*}$ invariant. Given an arbitrary choice of basis $\{\ket{j}_{A_{1}}\}$ for subsystem $A$, we choose 
  $\ket{j^{*}}_{B_{1}}$ as the basis for subsystem $B$, where $\ket{j^{*}}$ denotes the vector $\ket{j}$ with the corresponding complex conjugate amplitudes. For such a choice of basis the invariance of isotropic state guarantees the following
 \begin{equation}
 \label{singletermofmutualpredictability}
{_{B_{1}}}\langle j|_{A_{1}}\langle j|\rho_{I}|j\rangle_{B_{1}}| j\rangle_{B_{1}} = \frac{x}{d}+\frac{(1-x)}{d^{2}}.
 \end{equation}
Considering $O_{A_{1}}$ and $O_{A_{1}}^{*}$ to be the observables whose eigenvectors are $\{\ket{j}_{A_{1}}\}$ and $\{\ket{j^{*}}_{A_{1}}\}$, respectively. Then the mutual predictability is given as
 \begin{equation}
 \label{mutualpredictabilityiso}
C_{A_{1}B_{1}}=x+\frac{(1-x)}{d}.
 \end{equation}
For a set of choice of mutually unbiased observables $\{O_{A_{k}}\}$ and $\{O_{B_{k}}^{*}\}$ we then have
\begin{equation}
\label{summutualpredictabilityisotropic}
\sum_{k=1}^{m}C_{A_{k}B_{k}}=m\left(x+\frac{(1-x)}{d}\right).
\end{equation}
The R.H.S of Eq.~(\ref{summutualpredictabilityisotropic}) exceeds the upper bound for separability given by Eq.~(\ref{mutualpredictabilityiso}) if $x>\frac{1}{m}$. Thus, if there exists $m=(d+1)$ mutual unbiased observables then the mutual predictability is necessary and sufficient to detect entanglement of isotropic states. For $m=(d+1)$ and $x>\frac{1}{m}=\frac{1}{d+1}$, we have 
$$\sum_{i=1}^{d+1}C_{i}>2.$$
Using Eq.$(\ref{visibilitypredictability})$ for $m=(d+1)$, we then obtain
$$\sum_{i=1}^{d+1}V_{i}<(d-3),$$
thus, violating Eq.~(\ref{completevisibilityinequality}) for entangled isotropic state. Therefore, by suitable choice of unitary as given in Eq.~(\ref{equ:unitary_for_mutually_unbiased_observables}) one can detect whether the state is entangled or not.

\subsection{Determination of Entanglement witness from interferometric set-up}\label{subB}

For multipartite systems, there are entanglement witnesses which can
detect quantum entanglement~\cite{CG14}. These witness operators have positive average values on all separable states and negative on some entangled states.
\begin{definition}
A state $\rho$ is entangled if and only if there exists a Hermitian operator $W$ such that $\tr(W\rho)<0$ and for any separable state $\sigma$ we have $\tr(W\sigma)\geq 0$. The operator $W$ is known as entanglement witness.
\end{definition}
Let us consider the following entanglement witness in $\mathcal{H}_{A}\otimes\mathcal{H}_{B}$ with $\mathrm{dim}(\mathcal{H}_{A})=\mathrm{dim}(\mathcal{H}_{B}) = d $
\begin{equation}
\label{flip}
F_{AB} = \sum_{i,j}\ketbra{i}{j}\otimes\ketbra{j}{i}.
\end{equation}
We can show that $F_{AB}$ acts as entanglement witness for the Werner state defined as follows
\begin{equation}
\label{werner}
\rho_{w}= x Q_{s}+(1-x)Q_{a},
\end{equation}
where
\begin{align}
\label{symmetric}
Q_{s} & =\frac{2}{d(d+1)}\left(\mathbb{I}_{A}\otimes\mathbb{I}_{B}+F_{AB}\right)\\
\label{asymmetric}
Q_{a} & = \frac{2}{d(d-1)}\left(\mathbb{I}_{A}\otimes\mathbb{I}_{B}-F_{AB}\right).
\end{align}
It is well known that the state (\ref{werner}) is separable if and only if $ x \geq \frac{1}{2}$. From Eqs.~(\ref{flip}) and (\ref{werner}) we can obtain
\begin{equation}
\tr(F_{AB}\rho_{w})=2(2x-1).
\end{equation}
Thus, $\tr(F_{AB}\rho_{w})<0$ if and only if $x <\frac{1}{2}$ i.e., $\rho_{w}$ is entangled.\\

We can obtain the expectation value of an witness operator by considering $U=\exp[i\theta W]$, where $\theta$ is a real parameter. Let us apply the above unitary in one arm of the interferometer for an infinitesimal angle $\theta$. In this limit one can write $U\simeq (\mathbb{I}+i\theta W)$. One can then obtain 
\begin{align}
\label{wit1}
\tr(U\rho) & \simeq (1+i\theta\tr(W\rho),\\
\label{wit2}
& \simeq \exp[+i\theta \tr(W\rho)].
\end{align}

From Eq.~\eqref{wit2} we can infer that the phase of the intensity given by Eq.~(\ref{intensity}) is changed by an amount proportional to $\tr(W\rho)$. If $\tr(W\rho)<0$, then the state is entangled and the phase in Eq.~(\ref{intensity}) changed from $\phi$ to $\phi-(\theta\tr(W\rho))$. On the other hand, if $\tr(W\rho)>0 $, then the phase changed from $\phi$ to $\phi+(\theta\tr(W\rho))$. Thus, from observing the change in phase, we can determine whether the quantum state is entangled.

Note that the above formalism of detecting entanglement is limited to small values of $\theta$. However, we can go beyond it for certain types of witness operator. To see this, consider the SWAP operator defined in Eq.~(\ref{flip}). Note that $\tr(F_{AB})=d$, $F^{\dagger}_{AB}=F_{AB}$ and $F^{2}=\mathbb{I}_{A}\otimes\mathbb{I}_{B}$, where $\mathbb{I}_{A}$ and  $\mathbb{I}_{B}$ are identity operators associated with $\mathcal{H}_{A}$ and $\mathcal{H}_{B}$ respectively. Thus, $F_{AB}$ is Hermitian as well as unitary operator. Use of SWAP operator to detect entanglement in an interferometric set-up was proposed in Ref. \cite{Horodecki2002}. It was shown that an unitary involving SWAP operation acting on the two copies of the state shows the presence of entanglement in the interferometric visibility for pure state. On the other hand, we consider the capability of witnessing entanglement from the SWAP operator and use it as an example where we can go beyond the small values of $\theta$ in Eqs.~(\ref{wit1}) and (\ref{wit2}). To determine expectation value of $F_{AB}$ from an unitary operation let us take our unitary operator $U$ to be $F_{AB}$. We then have
\begin{equation}
\label{evolvedwerner}
\tr(U\rho_{w}) =\tr(F\rho_{w})=2(2x-1).
\end{equation}
If $\rho_{w}$ is entangled then the right hand side of Eq.~(\ref{evolvedwerner}) is negative otherwise it is positive. In other words
\begin{equation}
\label{entangledwerner}
\tr(U\rho_{w})=\begin{cases}
2(2x-1)e^{i\pi}  & \text{when $\rho_{w}$ is entangled}\\
2(2x-1)  & \text{when $\rho_{w}$ is separable}.
 \end{cases}
\end{equation}

Thus, if $\rho_{w}$ is entangled then the phase in Eq.~(\ref{intensity}) changes from $\phi$ to $\phi + \pi $. If $\rho_{w}$ is separable then the phase in Eq.~(\ref{intensity}) does not change. Thus, our proposal provides a phase sensitive method to detect entangled and separable states for the Werner state. In future, it will be worth exploring how to design a phase-sensitive entanglement meter which measures entanglement based on the above result.


\section{Conclusions}\label{conclusion}
To summarise, we have proposed several methods to efficiently certify and quantify entanglement of composite systems for qubits as well as higher dimensional systems using Mach-Zehnder interferometer. 
We have shown that having access to single copy of one of the subsystem and a suitable oracle, we can measure the linear entanglement entropy for any bipartite pure states in any finite dimension. In particular, our result shows that for any two qubit pure bipartite state, the interference visibility is a direct measure of entanglement. For arbitrary bipartite mixed states, using convex roof construction, we provide an upper bound to the linear entropy of entanglement. We have also proposed another interferometric scheme to measure the negativity of any pure bipartite state and some mixed states from the visibility. In addition to measuring entanglement content, we have shown how to detect entangled states using the interference visibility and phase shift.
Furthermore, we have proposed how to measure mutually predictability experimentally from the intensity patterns of the interferometric set-up without having to resort to local measurements of mutually unbiased basis. 
Towards the end, we have proposed how to measure the average of the witness operator in Mach-Zehnder interferometer and argued that the phase shift in the interference pattern can be a signature of entangled or 
separable nature of the input state. Thus, our proposal can have wide variety of applications in detection and quantification of entanglement in pure as well as mixed bipartite states. 
In addition, results presented in this paper can have interesting applications in design of entanglement meter where one can have a portable device to measure the entanglement content of bipartite states. 
The proposal can also be exploited to design phase-sensitive entanglement meter.
In future, it will be worth 
generalising these results for multipartite systems. We believe that our proposal can be experimentally tested with the existing technology.
\begin{acknowledgements}
SK acknowledges DST-QuEST fellowship. VP acknowledges Infosys Grant. AKP acknowledges support of the J.C. Bose Fellowship from the Department of Science and Technology (DST), India under Grant No.~JCB/2018/000038 (2019–2024).
\end{acknowledgements}
\bibliography{ref.bib} 
\end{document}